\documentclass[%
 reprint,
superscriptaddress,
 amsmath,amssymb,
 aps,
 pra,
dvipsnames,
twocolumn
]{revtex4-1}
\usepackage[main=english,german,ngerman]{babel}

\usepackage{lipsum}
\usepackage{xcolor}
\usepackage{graphicx}%
\usepackage{dcolumn}%
\usepackage{bm}%
\usepackage{siunitx}
\usepackage{dsfont}
\usepackage{tikz}
\usepackage{tikz-3dplot}
\usepackage{hyperref}
\usepackage{physics}
\usepackage{bbold}
\usepackage{float}
\usepackage{hyphenat}
\newcommand\figref{Fig.~\ref}

\usepackage[utf8]{inputenc}
\newcommand{\angstrom}{\mbox{\normalfont\AA}}
\usepackage{fixltx2e}

\usepackage[toc,page]{appendix}

\usepackage{tikz}
\usetikzlibrary{quantikz}

\usepackage{subfig}

\begin{document}

\title{Quantum neural networks force fields generation}

\author{Oriel Kiss}
\affiliation{%
IBM Quantum, IBM Research – Z\"urich, 8803 R\"uschlikon, Switzerland
 }%
\affiliation{Institute for Theoretical Physics, ETH Z\"urich - 8093 Z\"urich, Switzerland}
\affiliation{OpenLab, CERN - 1211 Geneva, Switzerland}
\author{Francesco Tacchino}
\affiliation{%
IBM Quantum, IBM Research – Z\"urich, 8803 R\"uschlikon, Switzerland
 }%
\author{Sofia Vallecorsa}
 \affiliation{OpenLab, CERN - 1211 Geneva, Switzerland}
\author{Ivano Tavernelli}
\affiliation{%
IBM Quantum, IBM Research – Z\"urich, 8803 R\"uschlikon, Switzerland
 }%
 \email{ita@zurich.ibm.com}

\date{\today}

\begin{abstract}

Accurate molecular force fields are of paramount importance for the efficient implementation of molecular dynamics techniques at large scales. In the last decade, machine learning methods have demonstrated impressive performances in predicting accurate values for energy and forces when trained on finite size ensembles generated with \textit{ab initio} techniques. At the same time, quantum computers have recently started to offer new viable computational paradigms to tackle such problems. On the one hand, quantum algorithms may notably be used to extend the reach of electronic structure calculations. On the other hand, quantum machine learning is also emerging as an alternative and promising path to quantum advantage. Here we follow this second route and establish a direct connection between classical and quantum solutions for learning neural network potentials. To this end, we design a quantum neural network architecture and apply it successfully to different molecules of growing complexity. The quantum models exhibit larger effective dimension with respect to classical counterparts and can reach competitive performances, thus pointing towards potential quantum advantages in natural science applications via quantum machine learning.

\end{abstract}

\maketitle

\section{Introduction}
Since more than half a century, atomistic simulations represent one of the sharpest tools available for scientific investigation %
in a wide range of research fields, such as chemistry, materials science and biology~\cite{MD_origins_1959,mccammon_dynamics_1977}. To perform a Molecular Dynamics (MD) calculation, in which the classical equations of motion are numerically integrated for each atom in the system under study, an accurate knowledge of potential energy surfaces (PES) and local forces is required. This information, originating from the quantum mechanical behaviour of electrons and nuclei, could in principle be deducted from the exact solution of the Schr\"{o}dinger equation. However, the complexity of such task makes it impractical beyond a few small-scale paradigmatic examples. In a delicate balance between performance and accuracy, approximate solution methods such as 
density functional theory (DFT) have therefore been proposed, leading to the family of so called \textit{ab initio} MD techniques~\cite{car_parrinello_1985,MarxHutterMD}. These precise yet computationally demanding strategies can be applied up to medium-sized systems, while larger problems may only be tackled, typically at a lower accuracy, with empirical force fields (FFs)~\cite{Unke_2020}. In fact, MD runs require on-the-fly computations of energy and forces at each time step and for each configuration visited by the system during its evolution: therefore, the use of simple parametrized functional potentials (e.g., the FFs) that can be evaluated in a fraction of the time required by actual quantum mechanical calculations is the only viable strategy when thousands of atoms are involved.

Recently, machine learning (ML) has emerged as a new technological paradigm offering promising and effective solutions for physics and chemistry~\cite{RevModPhysMLcarleo,ferminet}. In the context of MD simulations, a pioneering approach to ML-powered force fields was proposed by Beheler and Parrinello using neural networks~\cite{Parinnello}. The original idea has later been refined and extended~\cite{review_behler,behler_symmetry_2011,Rupp_PRL_2012,Morawietz8368,gastegger_wacsf_2018,symmetry,Unke_review_2020,cheng_evidence_2020}, also promoting the development of specific software libraries~\cite{lammps,Singraber_JCTC_2019}. The fundamental insight behind the so called neural network potentials (NNPs) is two-fold: first, they incorporate the idea that large performance gains can be achieved by directly modelling some form of functional relationship between structure (i.e., atomic positions) and properties of interest (e.g., energies), essentially bypassing the explicit solution of the underlying quantum mechanical problem. Second, NNPs typically enjoy the generalization capabilities of ML models, maintaining extremely good accuracy even on previously unseen configurations when trained on data sets constructed with ab initio methods.

Despite the success of classical ML techniques in the realm of atomic and molecular dynamical processes, the quantum mechanical character of the fundamental laws governing such phenomena immediately leads to the question whether quantum machine learning (QML) methods could provide further significant advantages~\cite{Biamonte2017QML,RevModPhysMLcarleo,huang2021quantum}. Indeed, also thanks to the high practical relevance of the problem, the learning and generation of molecular force fields may constitute a very natural and appealing playground in which QML could be tested and compared with state-of-the-art classical counterparts. An important aspect of such comparison lies in the fact that the properties to be learned, namely the relationship between configurations, energies and forces, are generally hard to be derived directly from first principles due to their quantum mechanical origin. At the same time, it has recently been suggested that information theoretical complexity considerations are strongly affected by the availability of training data even when quantum systems are involved~\cite{huang_power_2021}, with classical ML methods showing competitive performances, e.g., in predicting non-trivial many-body properties~\cite{huang2021provably}. Several important questions may therefore be addressed, from an overall assessment of the capabilities of QML protocols for PES and force fields reconstruction to systematic tests of classical versus quantum techniques for learning specific quantum mechanical properties through data.

In this work, we take a first step in such direction by establishing a direct connection between QML and NNPs. In particular, we demonstrate how quantum neural networks (QNNs) can be employed, in combination with classical data sets, as trainable models for the prediction of energies and forces in molecular systems. Although many different realizations of QNNs and quantum perceptrons are known in the literature~\cite{Schuld2015Simulating,cao2017quantum, Torrontegui2019Perceptron,Tacchino_2019,cong_quantum_2019,Tacchino_2020,Bondarenko2020TrainingDeep,Mangini_2020,Mangini_2021}, here we will focus on variational parametrized quantum circuits~\cite{Mitarai2018Learning,Benedetti2019Parametrized,SchuldPRAcircuitcentric2020,cerezo2020variational,tacchino_ieee_2021}, which offer the greater flexibility for near term applications. It is worth mentioning that some instances of QNNs, such as the ones that we will implement in the following, are known to exhibit greater power, as measured by the effective dimension in model space, compared to their classical equivalents~\cite{abbas_power_2021} -- a fact that places them among the most promising candidates in the quest for quantum advantage in machine learning. We also notice that, while variational quantum algorithms are often employed for the direct \textit{ab initio} computation of Hamiltonian spectra~\cite{peruzzo_variational_2014,vqe1} and corresponding forces~\cite{Sokolov_2021}, here we take a rather different approach~\cite{Kais2020Entropy}, using quantum resources to learn an implicit mapping between atomic coordinates, energy and forces, without any explicit solution of the quantum mechanical problem itself. %

\section{Model and methods}

\subsection{Quantum neural networks}

We adopt a supervised learning approach with training sets of the form $\mathcal{A} = \{(\vec{C}_\alpha,E_\alpha,\vec{F}_\alpha)\}$, namely collections of $n$-atom molecular configurations $\vec{C}_\alpha = (x_{a_1}^\alpha,y_{a_1}^\alpha,z_{a_1}^\alpha,\ldots,x_{a_n}^\alpha,y_{a_n}^\alpha,z_{a_n}^\alpha)$ with associated total energies $E_\alpha$ and forces $\vec{F}_\alpha = (F_{a_1,x}^\alpha,F_{a_1,y}^\alpha,F_{a_1,z}^\alpha,\ldots,F_{a_n,x}^\alpha,F_{a_n,y}^\alpha,F_{a_n,z}^\alpha)$. Here, the index $\alpha$ runs over the different elements of the training set, $c_{a_i}^\alpha$ for $c=x,y,z$ is the Cartesian coordinate of the $i$-th atom and 
\begin{equation}
    F_{a_i,c}^\alpha = - \frac{\partial E_\alpha}{\partial c^{\alpha}_{a_i}}
    \label{eq_force}
\end{equation}
is the force acting on atom $a_i$ along the direction $c$. We will also denote with $|\mathcal{A}|$ the number of samples contained in $\mathcal{A}$.

\begin{figure}
    \centering
    \includegraphics[width=\columnwidth]{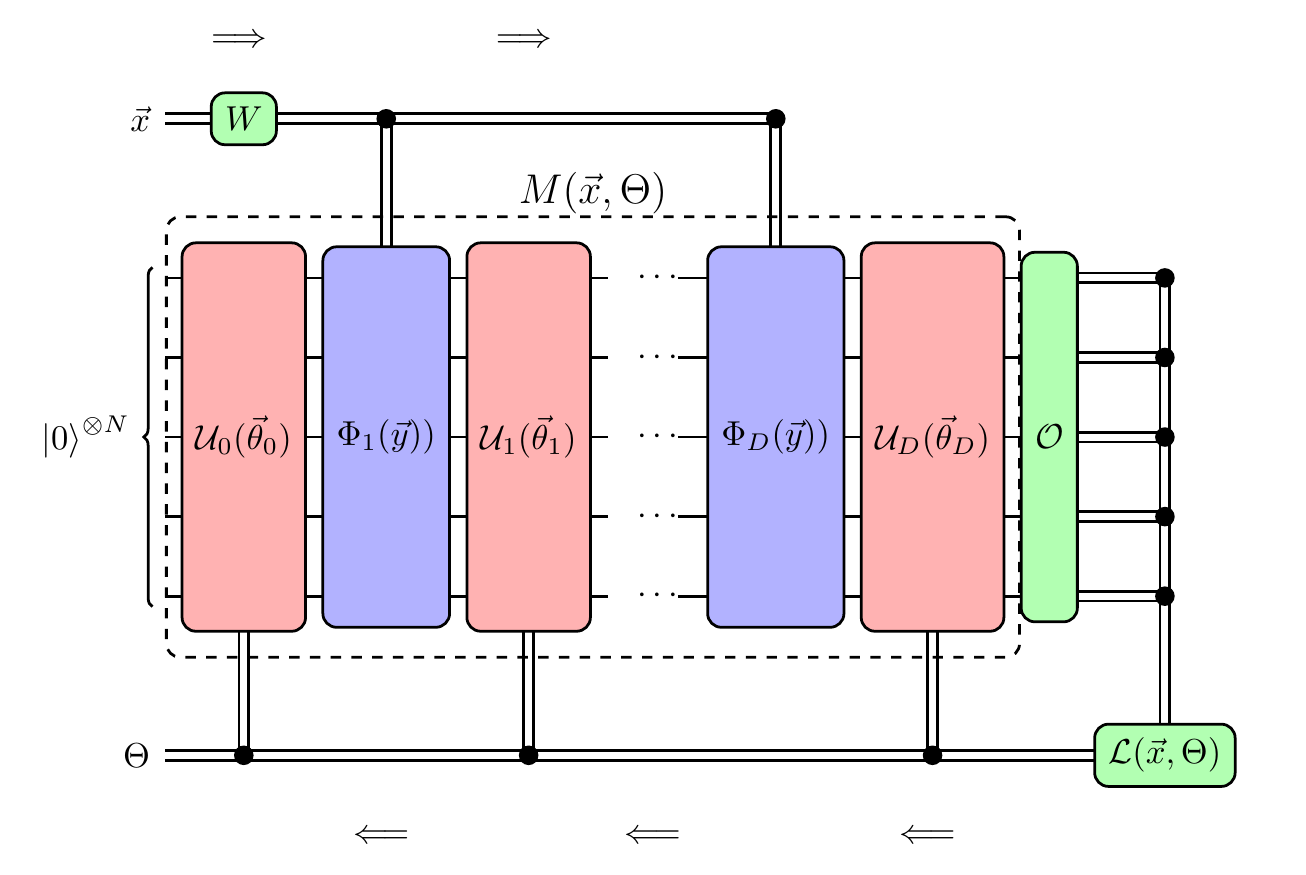}
    \caption{Quantum neural network model. The input data $\vec{x}$ are preprocessed with the classical (green) function $W$ and encoded with the map $\Phi$ (blue). The variational layers (red) contain free parameters that are optimized to minimize the loss function $\mathcal{L}(\vec{x},\Theta)$. $\mathcal{O}$ represents the computation of the expectation value of the observable through a quantum measurement.}
    \label{fig:qnn}
\end{figure}

The training data, derived from classical \textit{ab initio} methods that will be specified in the following, are used to optimize the predictions made by quantum models, specifically quantum neural networks. These are based on the general notion of parametrized quantum circuits (PQCs)~\cite{Benedetti2019Parametrized,cerezo2020variational,Mangini_2021}, consisting of a reference initial state $\ket{0}$, an output observable $\mathcal{O}$ and a model unitary $M(\vec{x},\Theta)$, depending on both some input data $\vec{x}$ and a set of trainable parameters $\Theta=\{\vec{\theta}_0,\ldots,\vec{\theta}_D\}$. 
The function expressed by a QNN takes the general form
\begin{equation}
    f_\Theta(\vec{x}) = \bra{0}M(\vec{x},\Theta)^\dagger \mathcal{O} M(\vec{x},\Theta)\ket{0}
    \label{eq_def_qnn}
\end{equation}
and can be represented schematically as shown in Fig.~\ref{fig:qnn}. In practical realizations, we allow for an additional classical preprocessing map $\vec{y} = W(\vec{x})$, which in our specific context can serve the purpose of converting between Cartesian coordinates and more suitable molecular descriptors, as well as enhancing the effective nonlinearity of the model~\cite{Mitarai2018Learning}. 

It is also worth noticing that, contrary to the usual feed-forward scheme of classical neural networks, QNNs are quite easily designed and interpreted in the form of re-uploading circuits~\cite{Qencoder,reuploaded}, where trainable layers $\mathcal{U}_\ell(\vec{\theta}_\ell)$, red in Fig.~\ref{fig:qnn}, are alternated with data encoding ones $\Phi_\ell(\vec{y})$, blue in Fig.~\ref{fig:qnn}, with the same classical input values appearing multiple times.

In fact, it has been shown that the re-uploading mechanism makes QNNs universal functional approximators~\cite{Vidal2020,Qencoder,reuploaded}: more specifically, any QNN output function (Eq.~\ref{eq_def_qnn}) can be recast into a truncated Fourier series with a set of available independent frequencies determined by the eigenvalues of the encoding map and growing with the number of re-uploading steps. While recent results suggest that, by expanding the size of the quantum registers, these models can actually be mapped back on simpler sequential ones in which all input operations appear at the beginning~\cite{jerbi2021quantum}, one can still profit from the insights offered by the re-uploading picture as a guide for intuition in the actual design of application-specific QNNs. As an example, theory suggests that, by adjusting the number of input layers, the richness of the Fourier spectrum can be systematically increased.

The structure of QNNs is completed by choosing a physical observable $\mathcal{O}$ and a suitable loss function $\mathcal{L}$, whose minimization drives the update of the trainable parameters via a classical optimization routine. In the following, we will make the simple choice
\begin{equation}
    \mathcal{O}=\sigma_z^1.
    \label{eq_observable}
\end{equation}
namely we will take the expectation value of the Pauli-$z$ operator on the first qubit to contruct the network output. Moreover, we will use a quadratic Mean Square Error (MSE) loss function 
\begin{equation}
    \mathcal{L}_\chi(\mathcal{A},\Theta) = \text{MSE(Energy)} + \chi\cdot \text{MSE(Forces)}.
    \label{eq_loss}
\end{equation}
with the hyperparameter $\chi$ weighting the contribution of energy and forces~\cite{Gastegger2020}. In practice, we directly associate the output of the QNN with the energy potential surface by defining
\begin{equation}
    \text{MSE(Energy)} = \frac{1}{|\mathcal{A}|}\sum_{\alpha \in \mathcal{A}} \Big(f_\Theta(\vec{C}_\alpha)-E_\alpha\Big)^2.
    \label{eq:MSE_energy}
\end{equation}
In parallel, consistent predictions for the forces are obtained by taking the derivative of the quantum circuit output with respect to the molecular coordinates:
\begin{equation}
    \text{MSE(Forces)} = \frac{1}{3n\cdot|\mathcal{A}|}\sum_{\alpha \in \mathcal{A}} \Big\Vert \grad_c f_\Theta(\vec{C}_\alpha)-\vec{F}_\alpha\Big\Vert^2.
    \label{eq:MSE_forces}
\end{equation}
We leave for future works the investigation of alternative strategies, such as the use of two independent quantum circuits for the separate learning of energies and forces.

\subsection{Encoding layers}\label{sec:feature_map}

As mentioned in the previous section, the encoding operations $\Phi_\ell(\vec{y})$ are used to input the molecular configurations in the QNN model. Here, we assume for simplicity that these encoding unitaries are identical across different layers or re-uploading stages, i.e., $\Phi_\ell(\vec{y})\equiv \Phi(\vec{y})$. Moreover, let us denote by $N$ the linear dimension of the feature vector $\vec{y}$, where in general $N$ can differ from $3n$ due to the preprocessing stage. Following standard practices~\cite{Qfeature}, we will use $N$ qubits to encode and manipulate $N$-dimensional features $\vec{y}=(y_1,\ldots,y_j,\ldots,y_N)$.

The quantum feature map $\Phi(\vec{y})$ is then constructed according to the expression
\begin{equation}
    \Phi(\vec{y}) = \mathcal{E}(\vec{y})\mathcal{S}(\vec{y})
\end{equation}
where $\mathcal{S}(\vec{y})$ is a collection of single-qubit Pauli-$y$ rotations
\begin{equation}
    \mathcal{S}(\vec{y}) = \prod_{j=1}^N \exp\left(-i\frac{\sigma_y^{(j)}}{2}y_j\right)
\end{equation}
and $\mathcal{E}(\vec{y})$ is an entangling operation of the form
\begin{equation}
    \mathcal{E}(\vec{y}) = \prod_{(j,k)\in \mathcal{P}} \exp\left(-i\sigma_z^{(j)}\sigma_z^{(k)}y_jy_k\right).
    \label{eq:ZZ_feature_map}
\end{equation}
The latter directly resembles the so called ZZ feature map originally introduced in Ref.~\cite{Qfeature}. The set of qubit pairs $\mathcal{P}$ can be chosen in different ways, balancing ease of implementation on physical architectures with functional expressivity. Standard examples include \textit{linear} entanglement
\begin{equation}
    \mathcal{P}_{\text{linear}} = \{(j,j+1)\, | \, j = 1,\ldots,N-1\},
\end{equation}
\textit{circular} entanglement
\begin{equation}
    \mathcal{P}_{\text{circ}} = \{(j,j+1 \text{ mod } N) \, | \, j = 1,\ldots,N\},
\end{equation}
and \textit{full} entanglement
\begin{equation}
    \mathcal{P}_{\text{full}} = \{(j,k) \, | \, j = 1,\ldots,N-1, k > j\}.
\end{equation}
For further generality, we also define a natural extension of the ZZ feature map to degree $l$ interactions, adding factors of the form
\begin{equation}
    \exp\left(-i\prod_{k=1}^ly_{j_k}\sigma_z^{(j_k)}\right)
\end{equation}
where, in principle, the $l$ qubits $(j_1,\ldots,j_l)$ can be arbitrarily chosen among the $N$ available.

In Fig.~\ref{fig_feature} we explicitly show a 3-qubit example with a \textit{full} 2-qubit entangling map and an additional $l=3$ operation. All multiple qubit operations are already decomposed into a standard universal set made of single-qubit rotations and CNOTs~\cite{RevTacchino}, which is typical of superconducting quantum computing architectures. \\

Before moving to the description of the trainable part of the QNN models, let us also remark a few points about the classical preprocessing step $W$ (see Fig.~\ref{fig:qnn}). As suggested above, this classical manipulation is generally used to enhance the nonlinear behaviour of the network, for example by taking inverse trigonometric functions of the original inputs~\cite{Mitarai2018Learning}. At the same time, we can use this initial step to embed the relevant set of physical symmetries into the abstract representation of the target molecular systems seen by the QNN. This is known to be crucial already for classical ML methods, where the role of symmetry preserving features is played, e.g., by the so called symmetry functions~\cite{Parinnello,behler_symmetry_2011}. 

To limit the complexity of the quantum models, in the following we will use a set of internal coordinates, namely bond distances and angles, which by design respect translational and rotational symmetries. The integration of more advanced techniques, including fragmentation of large systems into local atomic environments, the use of other classes of molecular fingerprints and, possibly, the realization of symmetry adapted quantum circuits all represent natural future extensions of the present work.

We explicitly notice that the conversion between Cartesian and internal coordinates ($\vec{I}_\alpha = W(\vec{C}_\alpha)$) must be taken into account when computing the quantum forces predictions, as these are defined with respect to the former class (Eq.~\ref{eq_force}). As a result, Eqs.~\ref{eq:MSE_energy} and~\ref{eq:MSE_forces} are more properly rewritten as
\begin{equation}
    \text{MSE(Energy)} = \frac{1}{|\mathcal{A}|}\sum_{\alpha \in \mathcal{A}} \Big(f_\Theta\left(W(\vec{C}_\alpha)\right)-E_\alpha\Big)^2
    \label{eq:MSE_energy2}
\end{equation}
and 
\begin{equation}
    \text{MSE(Forces)} = \frac{1}{3n\cdot|\mathcal{A}|}\sum_{\alpha \in \mathcal{A}} \sum_{c,a_i}\Big( \grad_I f_\Theta(\vec{I}_\alpha)\cdot \frac{\partial W}{\partial c^{\alpha}_{a_i}}-F^\alpha_{a_i,c}\Big)^2
    \label{eq:MSE_forces2}.
\end{equation}

\tikzset{
     operator/.append style={fill=blue!30,rounded corners},
     my label/.append style={above right,xshift=0.3cm},
     phase label/.append style={label position=above}
    }

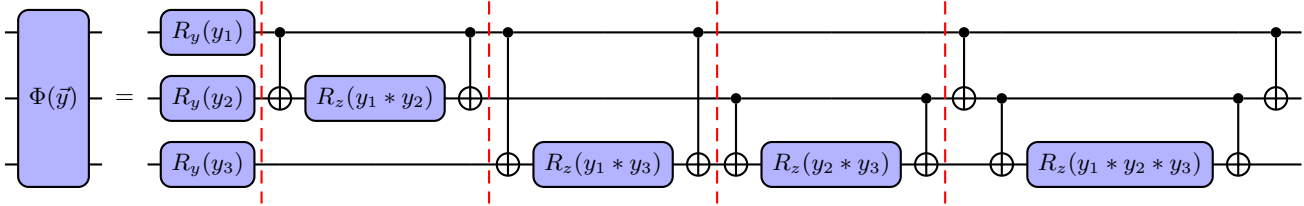
\begin{figure*}
\begin{center}
\begin{quantikz}[column sep=5pt, row sep={25pt,between origins}]
&\gate[3]{\Phi(\vec{y})}&\qw & &&\gate{R_y(y_1)}\slice{}&\ctrl{1} &\qw&\ctrl{1}\slice{}&\ctrl{2}&\qw&\ctrl{2}\slice{}&\qw&\qw&\qw\slice{}&\ctrl{1}&\qw&\qw&\qw&\ctrl{1}&\qw\\
&&\qw&=&&\gate{R_y(y_2)}&\targ{}&\gate{R_z(y_1*y_2)}&\targ{}&\qw&\qw&\qw&\ctrl{1}&\qw&\ctrl{1}&\targ{}&\ctrl{1}&\qw&\ctrl{1}&\targ{}&\qw&&\\
&&\qw&&&\gate{R_y(y_3)}&\qw&\qw&\qw&\targ{}&\gate{R_z(y_1*y_3)}&\targ{}&\targ{}&\gate{R_z(y_2*y_3)}&\targ{}&\qw&\targ{}&\gate{R_z(y_1*y_2*y_3)}&\targ{}&\qw&\qw&&
\end{quantikz}
\end{center}
\caption{Example of a \emph{full} encoding map with up to order $l=3$ interactions on $N=3$ qubits.}
\label{fig_feature}
\end{figure*}

\subsection{Trainable layers}

The parametrized operations $\mathcal{U}_\ell(\vec{\theta}_\ell)$ represent the adjustable components of the QNN model. If suitably optimized during the learning phase, these select the most appropriate mapping from inputs (molecular coordinates) to outputs (energy and forces).

The precise structure of the $\mathcal{U}_\ell(\vec{\theta}_\ell)$ can vary significantly across different models and applications, with several popular choices known in the literature. For our specific functional regression problem, we follow the steps of Mitarai et al.~\cite{Mitarai2018Learning} and make use of a physics-inspired ansatz which is known to generate highly entangled states. In particular, we consider the following fully connected transverse field Ising Hamiltonian
\begin{equation}
    H=\sum_{j=1}^N a_j \sigma_y^{(j)} + \sum_{j=1}^N\sum_{k=j}^NJ_{jk}\sigma_z^{(j)}\sigma_z^{(k)},
\end{equation}
and we use the induced time evolution operator $U(t) = e^{-iHt}$ as a template for the trainable layers. By using the approximate product formula
\begin{equation}
    e^{-itH} 
     \approx \prod_{j=1}^{N}e^{-ita_j\sigma_y^{(j)}}\cdot \prod_{j=1}^N \prod_{k=j}^N e^{-itJ_{jk}\sigma_z^{(j)}\sigma_z^{(k)}}
\end{equation}
and by replacing $a_jt$ and $J_{jk}t$ with free parameters, we obtain a unitary operation of the form
\begin{equation}
    \mathcal{U}_\ell(\vec{\theta}_\ell) = \prod_{j=1}^{N}e^{-i\theta_\ell^j\sigma_y^{(j)}}\cdot \prod_{j=1}^N \prod_{k=j}^N e^{-i\theta_\ell^{jk}\sigma_z^{(j)}\sigma_z^{(j)}},
\end{equation}
where now $\vec{\theta}_\ell$ is the collection of all $\theta_\ell^j$ and $\theta_\ell^{jk}$. It is easy to see that $\mathcal{U}_\ell(\vec{\theta}_\ell)$ can be implemented with single-qubit Pauli-$y$ rotations and two-qubit ZZ operations, similarly to what happens for the encoding map $\Phi$ defined in the previous section. We stress however that, while the encoding layers are identical across different reuploading stages, as they repeatedly input the same classical data $\vec{y}$, the trainable parameters $\vec{\theta}_\ell$ are allowed to vary across different layers $\ell = 0,\ldots,D$.

For $\ell = 0$, we make the special choice $\theta_\ell^{jk} = 0 \,\forall j,k$, namely we only use single-qubit rotations at the beginning of the computation (see Fig.~\ref{fig:qnn}). Moreover, we empirically find that including generalized $l$-qubit interactions up to $l=3$, i.e.~terms of the form $e^{-i\theta_{jkm}\sigma_z^{(j)}\sigma_z^{(k)}\sigma_z^{(m)}}$ improves the overall performances of the model at the cost of only a modest increase of circuit complexity. %

\subsection{Model training}

We train QNN models by minimising an average quadratic error that in principle contains both energy and forces labels, see Eq.~\eqref{eq_loss}. In most instances, we make use of an update rule which follows the negative gradient direction of the loss function
\begin{equation}
    \Theta^{t+1}=\Theta^t-\eta\nabla \mathcal{L}_\chi(\mathcal{A},\Theta^t),
    \label{eq_gd}
\end{equation}
where $\eta$ is the learning rate and $\Theta^t$ the set of trainable parameters at the optimization step $t$. For energy predictions derived from a quantum circuit, derivatives with respect to any given trainable parameter $\mu$ can be easily computed with the parameter shift rule~\cite{QG}
\begin{equation}
    \frac{\partial E_\alpha^\text{QNN}}{\partial \mu} = \frac{1}{2}\left[f_{\mu+\frac{\pi}{2};\Theta_\mu}(W(\vec{C}_\alpha))-f_{\mu-\frac{\pi}{2};\Theta_\mu}(W(\vec{C}_\alpha))\right],
    \label{eq_shift}
\end{equation}
where $E_\alpha^\text{QNN} = f_\Theta\left(W(\vec{C}_\alpha)\right)$ and $\Theta_\mu$ is the set of all trainable parameters without $\mu$. The corresponding result for the forces predictions is obtained with the iterative shift rule~\cite{hessian}, and schematically reads
\begin{align}
\begin{split}
\frac{\partial F_{a_i,c}^{\alpha,\text{QNN}}}{\partial \mu} &= -\frac{\partial^2 E^{\text{QNN}}_\alpha}{\partial c_{a_i} \partial \mu} = 
- \frac{1}{4}\sum_j\Big[f_{\mu+\frac{\pi}{2};\Theta_\mu}(\vec{I}_\alpha +\frac{\pi}{2}\vec{e}_j)\\
&-f_{\mu-\frac{\pi}{2};\Theta_\mu}(\vec{I}_\alpha +\frac{\pi}{2}\vec{e}_j) -f_{\mu+\frac{\pi}{2};\Theta_\mu}(\vec{I}_\alpha -\frac{\pi}{2}\vec{e}_j)\\
&+f_{\mu-\frac{\pi}{2};\Theta_\mu}(\vec{I}_\alpha -\frac{\pi}{2}\vec{e}_j) \Big] \cdot \frac{\partial I_j^\alpha}{\partial c_{a_i}},
\label{eq_it_shift}
\end{split}
\end{align}
with $(\vec{e}_j)_k=\delta_{jk}$ the standard basis vectors.

Notice that the parameter shift rule of Eq.~\eqref{eq_shift} can also be used to retrieve the actual QNN forces predictions, namely
\begin{equation}
    F^{\alpha,\text{QNN}}_{a_i,c} = -\grad_I f_\Theta(\vec{I}_\alpha)\cdot \frac{\partial W}{\partial c_{a_i}} = -\sum_j \frac{\partial f_\Theta(\vec{I}_\alpha)}{\partial I^\alpha_j}\frac{\partial I^\alpha_j}{\partial c_{a_i}} \, .
\end{equation}
In this case, factors of the form $\partial f/\partial I_j$ must be computed from the quantum circuit, while the chain rule factors taking into account the classical preprocessing are known analytically and are determined solely by the mapping function $W$.

At the beginning of training, all free parameters are initialized at zero, so that each hidden layer acts as the identity operator. This essentially follows the recommendations given in Ref.~\cite{initialization} to promote effective optimization steps in the early training phase.

\subsection{Effective dimension}

In the remaining part of this work, we will provide a series of examples to demonstrate how QNN models are able to learn and give consistent predictions of potential energy surfaces and force fields for individual molecules. The overall performances will be assessed primarily through the evaluation of the average root mean square error on suitable test sets. 

In addition to that, we will also make use of the concept of effective model dimension -- originally introduced in Ref.~\cite{abbas_power_2021} -- to inform the comparison with classical counterparts and to investigate potential advantages. As a brief summary, the effective dimension quantifies the capacity of both classical and quantum parametrized machine learning models~\cite{abbas2021effective} by measuring the portion of model space that they occupy. In other words, it estimates the capability of a model in covering the functional space defined by a particular model class by making a productive use of all its parameters, going beyond naive parameter-counting arguments. A high effective dimension is therefore related to a richer set of expressible functions and better trainability, as it can lead to a more favourable landscape for gradient based methods~\cite{abbas_power_2021}. Moreover, the effective dimension can sensibly bound the model generalization error~\cite{abbas_power_2021,abbas2021effective}.

For a given statistical model $y=f_\Theta(x)$, the effective dimension is defined as
\begin{equation}
    d_n(f_\Theta) = \frac{2\log{(\frac{1}{V_\Omega}\int_{\Omega}\sqrt{\det(\mathbb{1}+\frac{ n}{2\pi\log{n}}F(\Theta))}\,d\Theta)}}{\log{(\frac{n}{2\pi \log{n}}})}
\end{equation}
where $\Omega\subset \mathbb{R}^d$ is a $d$-dimensional parameter space of volume $V_\Omega$, $n\in\mathbb{N}$ is the number of data samples. Here, we have also introduced the Fisher information matrix
\begin{align}
    F &= \mathbb{E}[\nabla_\Theta \log{(p(x,y;\Theta)}\nabla_\Theta \log{(p(x,y;\Theta)}^T]\\
    &\approx \frac{1}{K}\sum_{k=1}^K\nabla_\Theta f_\Theta(x_k)\nabla_\Theta f_\Theta(x_k)^T,
\end{align}
with $p(x,y;\Theta)=p(y|x;\Theta)p(x)$ and $p(y|x;\Theta) = \exp(-\norm{y-f_\Theta(x)}^2/2)/\sqrt{2\pi}$ being the probability distribution mass of the model. The effective dimension is bounded by the rank of the Fisher information matrix and is usually normalized with the total number of parameters $d$.

\section{Results}

\subsection{Diatomic molecule: LiH}\label{sec:LiHqnn}

As a first proof-of-concept, we consider a single LiH molecule and we design a QNN model that learns its dissociation curve, and the corresponding force field, as a function of the bond length $r$.

The presence of a single internal coordinate, which is obtained from the Cartesian positions of the Li and H atoms with a mapping $W(\vec{C}) = |\vec{C}_\mathrm{Li}-\vec{C}_\mathrm{H}|$, makes the problem effectively one dimensional. We construct a classical dataset by numerically diagonalizing the second quantized Hamiltonian expressed in the STO3G basis set~\cite{vqe1} for different bond lengths $r$ in the range $[0.9, 4.5]$~\AA. Forces are computed via finite differences over the exact potential energy surface. 

To make better use of the Fourier series structure of the QNN, we make the exact potential energy surface symmetric by mirroring it around the $r = 4.5$ {\AA} and selecting the data set over the extended range, see Appendix~\ref{app:mirroring}. Furthermore, we use an additional preprocessing step -- on top of the Cartesian to internal coordinate transformation -- to construct 3-dimensional features from the original 1D problem. First, we scale all inputs in $[-1,1]$ with the \texttt{scikit\hyp learn} MinMaxScaler, then we apply the map
\begin{align}
\begin{split}
    W : [-1,1] &\rightarrow [-\pi,\pi]^3 \\
    r & \mapsto \begin{pmatrix} \pi r \\ \arcsin{(r)} \\ \arccos{(r)}\end{pmatrix}
    \end{split}
\end{align}

We employ a 3-qubit QNN model with $D=10$ trainable layers, alternated with the same amount of input stages. We use a \emph{full} feature map, as introduced in Sec.~\ref{sec:feature_map}, with $l=3$ (see also \figref{fig_feature}), and similar trainable layers with full entanglement and degree $l=3$ interactions. 

We benchmark our model against a fully connected classical neural network (NN), taking the 7 inputs encoded in the feature map, as in Fig. \ref{fig_feature}. The classical NN is composed of $5$ layers with $[7,4,5,2,1]$ units using the hyperbolic tangent activation function. Such network topology is chosen through a random search over all possible configurations with the same number of parameters (identical to the QNN), selecting the one that achieves the best validation loss. 

Both the classical and the quantum neural network models are trained with 4000 steps of the ADAM~\cite{adam} algorithm on $50$ data points using the $\mathcal{L}_0$ ($\chi = 0$) loss. We remark that the small size of the classical neural network makes it quite sensitive to parameters initialization, even when using the Xavier~\cite{xavier} scheme. Hence, the model underfits in $65\%$ of the cases and needs 1600 epochs to reach convergence. On the other hand, the quantum neural network appears to be robust against weights initialization and converges in 230 epochs.

\begin{figure}
    \centering
    \subfloat[\centering LiH Energy]{\includegraphics[width=8cm]{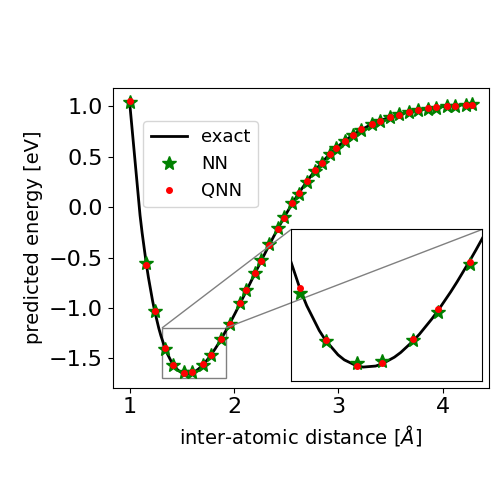}}%
    \qquad
    \subfloat[\centering LiH Force]{\includegraphics[width=8cm]{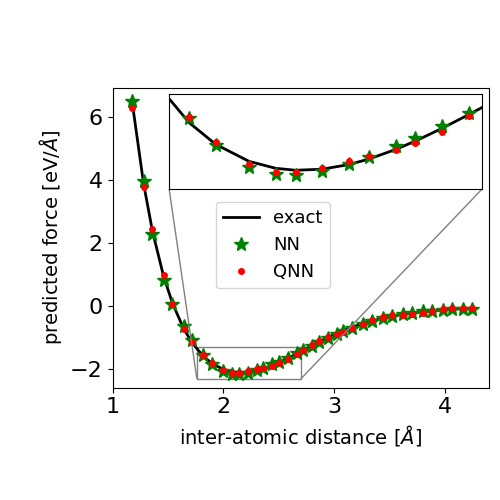} }%
    \caption{Prediction of the LiH energy (shifted by $-212.8$ eV)  (a) and force (b) as a function of the the inter\hyp atomic distance $r$. The exact solution (black line) is compared with the QNN (red dots) and the classical NN (green stars). Insets show an enlarged view of the energy minimum region.}
    \label{LiH_prediction}
\end{figure}

The validation root mean square errors (RMSE) for both models, computed on a test set with $120$ data points, are given in Table \ref{tab:liH_loss} and the respective predictions are shown in Fig.~\ref{LiH_prediction}. We notice that, in this setting, the QNN outperforms the best classical counterpart with the same number of parameters in terms of stability and quality of the predictions, and also exhibits a larger effective dimension. Although the classical model can match the QNN results in absolute terms when its size grows or if the forces are explicitly included in the loss function ($\chi \neq 0$), the present comparison, supported by the effective dimension analysis, certifies the competitiveness of the proposed quantum models. In Appendix~\ref{app:md} we also report an example of MD trajectories driven by the exact and learned force fields.
\begin{table}[]
    \centering
    \begin{tabular}{c|ccccc}
    %\hline 
    \textbf{LiH} & \textbf{RMSE(E)} & \textbf{RMSE(F)} & $\mathbf{d_{50}/d}$ & $\mathbf{d}$ & \textbf{Epochs}\\
    \hline 
      QNN   & $4\times10^{-3^{}} $& $0.05$& 0.9&73&230\\
      %\hline 
       NN  & $8\times 10^{-3}$&$0.6$&$0.38$&73&1600\\
       \hline 
    \end{tabular}
    \caption{Validation RMSE for LiH energy ([eV]) and forces ([eV/\AA]), together with the normalized effective dimension ($d_{50}/d$), the total number of parameters ($d$) and the number of training epochs for both the QNN and the classical NN models.}
    \label{tab:liH_loss}
\end{table}

\subsection{A single H$_2$O molecule}\label{sec:water} %

In the second example, we move towards the multi\hyp dimensional case by considering an individual H$_2$O molecule. For this more challenging test, we choose 3 internal coordinates, namely the two O-H bond lengths and the H-H planar angle, which we scale again in $[-1,1]$ and preprocess to create the following 3-dimensional feature vector
\begin{align}
\begin{split}
    W : [-1,1]^3 &\rightarrow [-\pi,\pi]^3 \\
    \begin{pmatrix} r_{\mathrm{OH},1} \\ r_{\mathrm{OH},2} \\ \phi_{\mathrm{HH}} \end{pmatrix} & \mapsto \begin{pmatrix} \arcsin{(r_{\mathrm{OH},1})} \\ \arcsin{(r_{\mathrm{OH},2})} \\ \arcsin{(\phi_{\mathrm{HH}})}\end{pmatrix}
    \end{split}
\end{align}
The classical configurations and the corresponding energy and forces labels, computed with density functional theory (DFT), are retrieved from Ref.~\cite{H2O_dataset}. To simplify the problem, we concentrate on configurations around the equilibrium position, discarding those with bond length outside $[1.6,2.1]$ \AA.

Our QNN model is constructed similarly to the LiH case, with three qubits and a depth of $D=12$ encoding and trainable layers. We compare again its results with those of a classical NN model with the same number of parameters and designed with the same conditions applied in the LiH case presented above. Furthermore, we also use as a reference the state-of-the-art specialized n2p2~\cite{n2p2} classical package, which makes full use of symmetry functions and of the sub-network fragmentation idea~\cite{Parinnello}, and from which we expect peak performance.

More specifically, the simplified classical NN model is a fully\hyp connected 6-layer network with $[7, 4, 6, 2, 2, 1]$ units  and hyperbolic tangent activation function. Instead, the n2p2 network uses three sub\hyp networks with two hidden layers, each with 15 units, and hyperbolic tangent activation. This model takes as input a set of 15 symmetry functions for the Oxygen atom and 20 for the two Hydrogen ones. The models are trained by minimising a $\mathcal{L}_1$ loss function on $300$ data points ($1000$ for n2p2). Notice that here we set $\chi=1$: indeed, as pointed out in Ref.~\cite{gradient_role}, it is important to incorporate the forces in the training if we wish to predict them accurately. However, this makes a full gradient descent computationally very intensive for the simulation of the quantum model, as it requires the calculation of the circuit Hessian matrix with recursive parameter shifts (see Eq. \ref{eq_it_shift}). For this reason, we choose the gradient free optimizer COBYLA for the present demonstration, while the classical models are trained with the ADAM method.

\begin{figure}
    \centering
    \subfloat[\centering H$_2$O Energy]{\includegraphics[width=8.5cm]{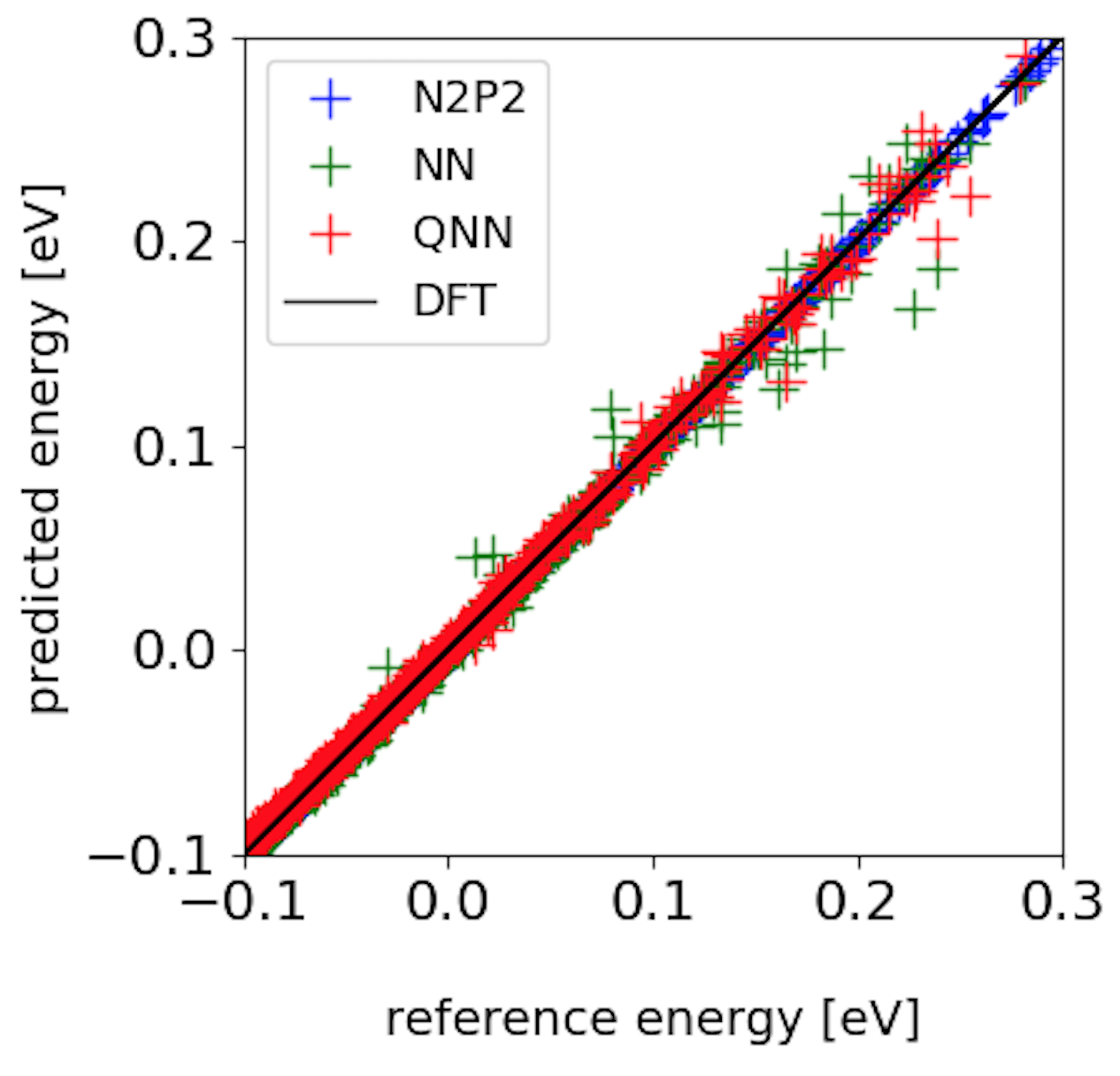}}%
    \qquad
    \subfloat[\centering H$_2$O Forces]{\includegraphics[width=8.5cm]{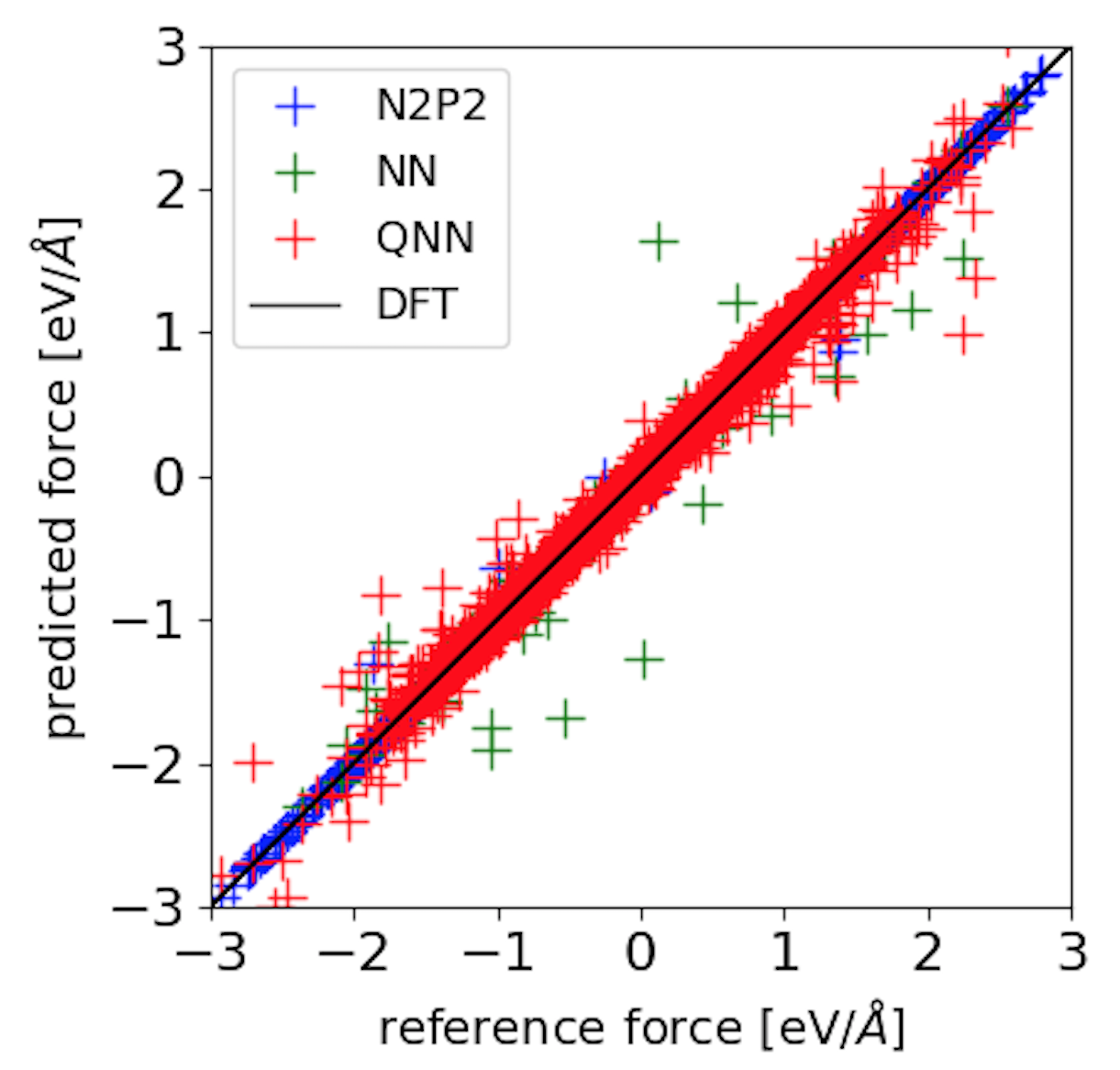} }%
    \caption{Predicted  H$_2$O energy (a) and forces (b) from the the QNN (red crosses), the generic NN (green crosses) and the n2p2 (blue crosses) models compared to reference DFT data points (black line).}
    \label{fig:H2Oprediction}
\end{figure}

The validation loss results (computed on a test set with $650$ data points) are presented in Table~\ref{tab:H2O_loss}, while Fig.~\ref{fig:H2Oprediction} shows the predicted energy and forces against the reference data points. The QNN is once again competitive with respect to classical counterparts, outperforming the non-specialized classical NN and reporting the largest normalized effective dimension.
\begin{table}[]
    \centering
    \begin{tabular}{c|cccc}
    %\hline 
    \textbf{H$_2$O} & \textbf{RMSE(E)} & \textbf{RMSE(F)} & $\mathbf{d_{300}/d}$ & $\mathbf{d}$\\
    \hline 
      QNN   &0.005 &0.06 &0.72&87\\
      %\hline 
        NN  & 0.006& 0.1 & 0.25&87\\
       %\hline 
         n2p2  & $7 \times 10^{-4}$&0.01&0.04&1642 \\
       \hline
    \end{tabular}
    \caption{Validation RMSE for H$_2$O energy ([eV]) and forces ([eV/\AA]), together with the normalized effective dimension ($d_{300}/d$) and the total number of parameters ($d$) for the QNN and the classical NN and n2p2 models.
    }
    \label{tab:H2O_loss}
\end{table}

\subsection{Umbrella motion of Hydronium}\label{sec:umbrella}

\begin{figure}
    \centering
    \subfloat[\centering H$_3$O Energy]{\includegraphics[width=8.5cm]{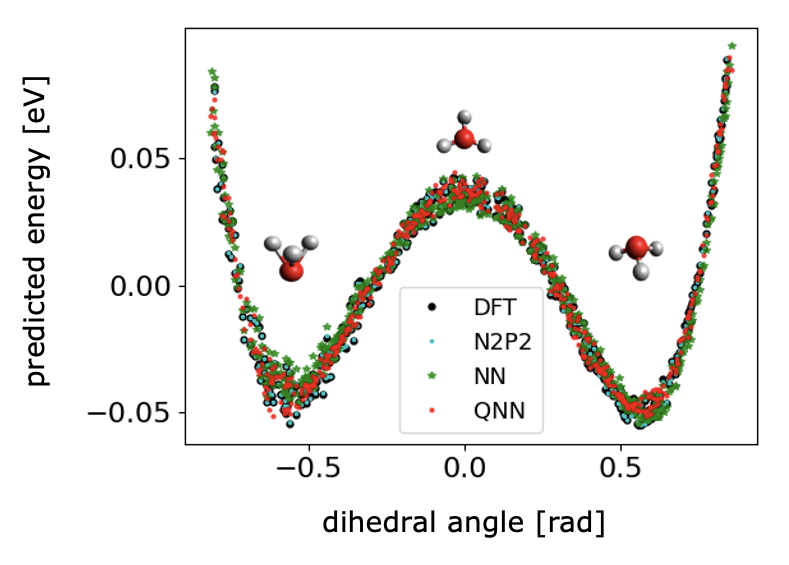}}%
    \qquad
    \subfloat[\centering H$_3$O Energy]{\includegraphics[width=8.5cm]{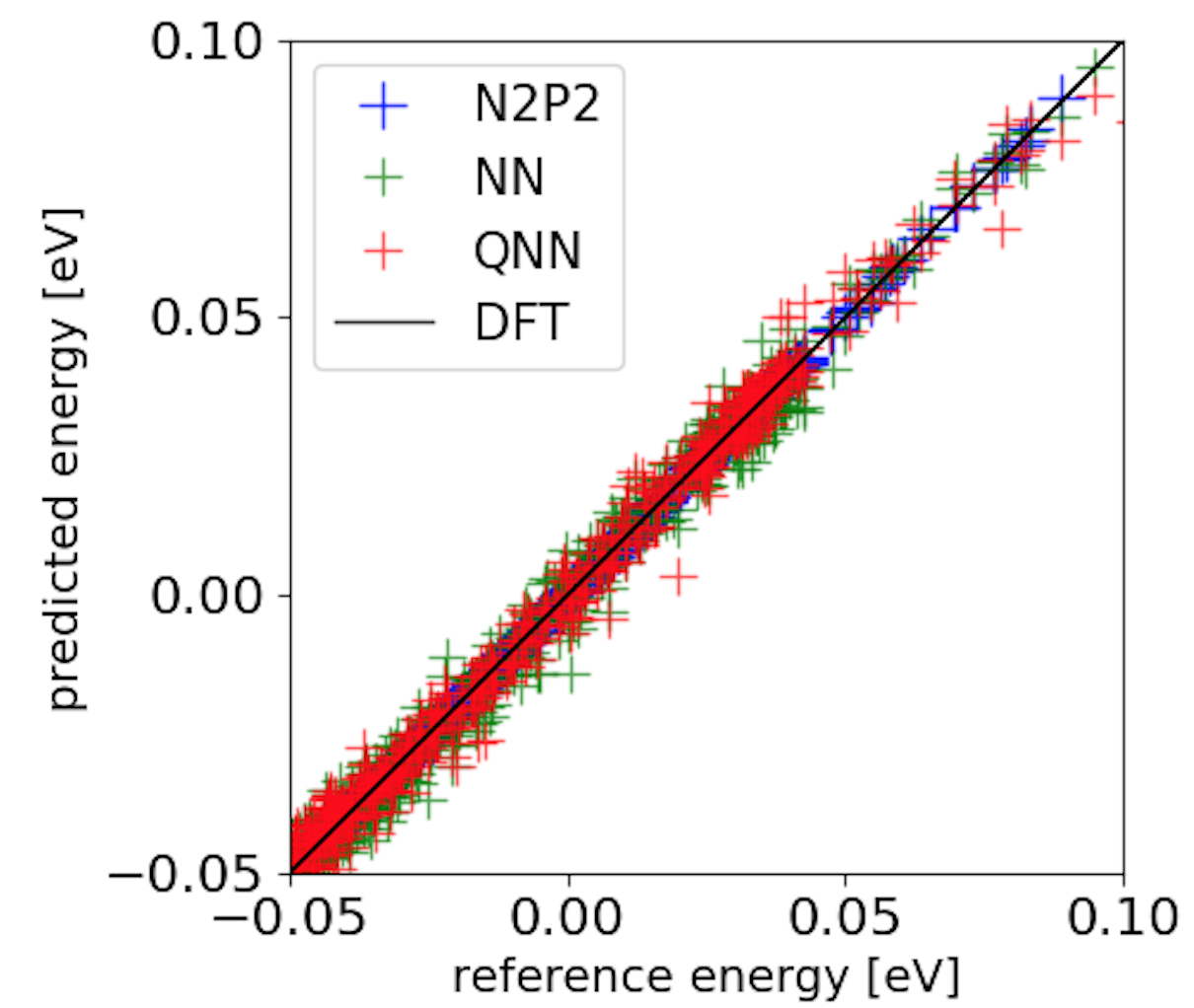} }%
    \caption{Predicted  H$_3$O$^+$ energy (shifted by -474.45 [eV]) from the QNN (red), the generic classica NN (green) and the n2p2 (blue) models as a function of the dihedral angle (a) and compared to the reference DFT energy (b). The molecule is represented at the two energy minima where the Oxygen atom sits above or below the plane formed by the Hydrogen ones, and at the saddle point where the Oxygen and Hydrogen atoms are co-planar.}
    \label{fig:H3O_prediction}
\end{figure}

In our last test, we consider a single Hydronium (H$_3$O$^+$) molecule. Following Ref.~\cite{H3O}, we prepare a training set by sampling configurations traversed along the inversion pathway shown in Fig.~\ref{fig:H3O_prediction} using DFT-based Molecular Dynamics (MD) simulations at 400 K and collecting the corresponding energies and forces. 
In order to sample the full profile along the dihedral angle `HHHO' from -0.78 to 0.78 rad, we also applied a dynamical constraint with increments of 0.005 rad at each MD time step.
All calculations were performed with the plane-wave (PW) code CPMD~\cite{CPMD} using unrestricted Kohn-Sham DFT with the PBE functional~\cite{Perdew_PRL1996}, a PW cutoff of 70 Ry, a cubic simulation box of edge 14 \AA, and Trouiller-Martins pseudo-potentials~\cite{trouiller1991}. For the MD, a time step $10$ a.u. ($0.242$ fs) was used.

We describe each configuration with 6 internal degrees of freedom, namely three O-H bond lengths, two O-H angles and the dihedral angle formed by the four atoms (see Appendix~\ref{app:dihedral} for the formal definitions). We also make use of the explicit formulas for derivatives with respect to Cartesian coordinates provided in Ref.~\cite{internal}. 

A 6-qubit QNN model is constructed with $D=10$ repetitions of the encoding map and the trainable layers, both with \textit{linear} entanglement and order $l=3$ interactions. The latter are also placed with linear couplings across all neighbouring 3-qubit groups. Due to the complexity of the system and the high computational cost, we only include energy labels in the training set, leaving a more refined study of the forces for future investigations. Hence, we train the model by minimizing a $\mathcal{L}_0$ loss on 500 data points (9000 for n2p2) with 5000 steps of the ADAM algorithm.

As in the previous examples, we compare the QNN model with a classical neural network, whose topology is optimized under the constraint that the number of trainable parameters be the same of the QNN, and with a n2p2-build model. In this case, the classical NN has [6, 14, 2, 1] units, while the n2p2 neural network is composed of 4 sub\hyp networks, each one with a structure similar to the case if the H$_2$O molecule in Sec.~\ref{sec:water}.

The RMSE results for a test set with 500 data points are reported in Tab.~\ref{tab:H3O_loss}, and energy predictions are also depicted in Fig.~\ref{fig:H3O_prediction}. Here, the QNN model is still competitive with respect to the generic classical NN, although both of then are clearly outperformed by the specialized and much larger n2p2 model. It is also worth noticing that forces predictions are much worse than energy ones, which confirms the necessity of including them in the loss function to achieve good results in non-trivial systems. 
\begin{table}[]
    \centering
    \begin{tabular}{c|cccc}
    \textbf{H$_3$O$^+$} & \textbf{RMSE(E)} & \textbf{RMSE(F)} & $\mathbf{d_{500}/d}$ & $\mathbf{d}$\\
    \hline 
      QNN &3.7$\times 10^{-3}$  & 0.26 & 0.67&135\\
        NN  &4.2$\times 10^{-3}$ & 0.207 & 0.19&135\\
         N2P2 &5$\times 10^{-4}$  & 0.19 & 0.03&2214\\
       \hline
    \end{tabular}
    \caption{Validation RMSE for H$_3$O$^+$ energy ([eV]) and forces ([eV/\AA]), together with the normalized effective dimension ($d_{500}/d$) and the total number of parameters ($d$) for the QNN and the classical NN and n2p2 models.}
    \label{tab:H3O_loss}
\end{table}

\section{Conclusions} % 

In this work we have successfully demonstrated the systematic application of quantum machine learning techniques, and specifically parametrized quantum neural networks, to the problem of learning molecular potential energy surfaces from classical \textit{ab initio} data sets and for the generation of molecular force fields. In all our numerical simulations, the proposed QNN models already achieved competitive performances with respect to comparable classical ones, reporting good prediction accuracy for several paradigmatic single-molecule examples. 

The present assessment naturally opens several questions and future research directions. On the one hand, the design of more specialized QNN architectures and molecular descriptors would allow a more refined and effective treatment of the problem, including the possibility of tackling bulk materials. For example, classical state-of-the-art implementations~\cite{n2p2} crucially benefit from the fragmentation of large systems in local environments and corresponding sub-networks~\cite{Parinnello}. Similar techniques could possibly be engineered for QNNs, e.g.~by resorting to more general perceptron-based models~\cite{Tacchino_2020,Bondarenko2020TrainingDeep}. These could explicitly be partitioned into local components and entangling connections between different sub-networks could eventually be realized. In addition to that, the development of efficient quantum gradient estimation protocols will be crucial for larger scale implementations of the proposed methods and for an adequate treatment of forces. It is also worth mentioning that another viable alternative may represented by quantum kernel methods~\cite{Qfeature}, whose classical counterparts are already extensively used in the context of machine learning potentials~\cite{review_behler}, for example in combination with Coulomb matrix descriptors~\cite{Rupp_PRL_2012}.

On the other hand, the most interesting open questions concern the potential quantum advantages brought by QML approaches. In this work, we focused on the problem of learning from classical data and we observed, following Ref.~\cite{abbas_power_2021}, that suitably designed QNN models exhibit a larger effective dimension than their classical equivalents. This in turn relates to stable and fast training capabilities, and points toward a more effective handling of larger systems. Indeed, large normalized effective dimensions signal, in the spirit of capacity measures, an effective use of the available model parameters, thus suggesting that quantum models could represent economic and manageable tools to tackle large molecular simulations. 

At the same time, this analysis does not yet take into account the role of overparametrization, which is known to contribute in a crucial way to the performances of classical neural networks. 
In fact, our numerical experiments confirm that large classical models, such as the ones employed in Sec.~\ref{sec:water}-\ref{sec:umbrella}, still achieve the best prediction and generalization accuracy. 
Interestingly, while the normalized effective dimension of those classical models is quite small, the absolute effective dimension is actually comparable to the one of their direct quantum counterpart, thus hinting at some form of `computational phase transition' in their behaviour~\cite{Belkin15849,nakkiran2019deep,larocca2021theory}. 
A few similar observations have already been made for quantum machine learning models~\cite{larocca2021theory}, and a thorough exploration of QNNs capabilities in such highly overparametrized regime -- including, among others, the question of whether this could be realized in a more effective way or with different qualitative behaviours compared to classical models -- represents an interesting open research direction in general. We leave a complete analysis of its application in the context of NNPs for future investigations.

On a larger perspective, one may also envisage the use of QML methods on quantum data~\cite{cong_quantum_2019,huang2021quantum} retrieved from experiments or quantum chemistry simulations, e.g., in the form of quantum wavefunctions generated through variational~\cite{vqe1,mcardle2020} or dynamical methods. The extraction of physical/chemical properties and the classification of materials directly at the quantum level of description could then likely represent one of the most advanced and exploratory efforts in the quest for quantum advantage with quantum machine learning.

\section*{Acknowledgements}
We thank Amira Abbas and Stefan W\"orner for insightful discussions. We acknowledge financial support from the Swiss National Science Foundation (SNF) through the grant No.~200021-179312. O.K.~acknowledges financial support through a scholarship from the Dominik and Patrick Gemperl\'e Foundation.

IBM, the IBM logo, and ibm.com are trademarks of International Business Machines Corp., registered in many jurisdictions worldwide. Other product and service names might be trademarks of IBM or other companies. The current list of IBM trademarks is available at \url{https://www.ibm.com/legal/copytrade}.

\bibliography{bibliography}

%

%\newpage

\appendix

\section{LiH -- Mirroring}\label{app:mirroring}

As we recalled in the main text, quantum neural network models introduced in Eq.~\ref{eq_def_qnn} can be expressed as partial Fourier Series~\cite{Qencoder} 
\begin{equation}
    f_\Theta(x)=\sum_{n\in \Omega} c_n e^{inx},
    \label{fourier_series}
\end{equation}
where $\Omega$ is the set of available frequencies and depends exclusively on the encoding map, while the coefficients $c_n$ are determined by the trainable unitary gates and the observable. QNNs of this form become universal functional approximators in the limit $|\Omega|\rightarrow \infty$. In practice, we make use of the Fourier series interpretation of QNNs and of Fourier analysis to guide model design, for example by changing model complexity (in our case essentially determined by the depth $D$ for fixed qubit number) to control the number of available frequencies. 

\begin{figure}
   \centering
    \includegraphics[width=8.5cm]{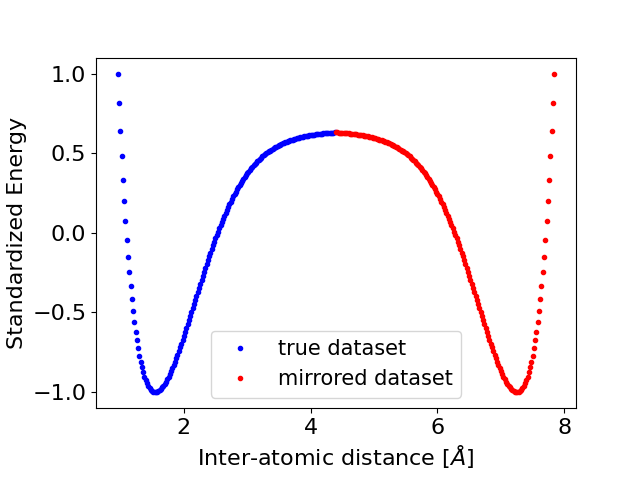}
    \caption{LiH data set (blue) and the corresponding mirrored extension (red).}
    \label{fig:mirror}
\end{figure}

In simple proof-of-principle experiments, such as LiH, we can also compare the model spectrum with the one of the data, although this is may not be possible in general. We also empirically find that better performances are obtained for periodic data sets, which intuitively correspond to finite frequencies spectra and hence reduce spurious oscillations in the final solution. We apply this intuition to the the 1-dimensional LiH case, whose PES can be artificially made periodic by mirroring around $r=2.5$ \AA, as shown in Fig.~\ref{fig:mirror}.

\begin{figure}
    \centering
    \subfloat[\centering Time evolution of the inter-atomic distance ]{\includegraphics[width =8.5cm]{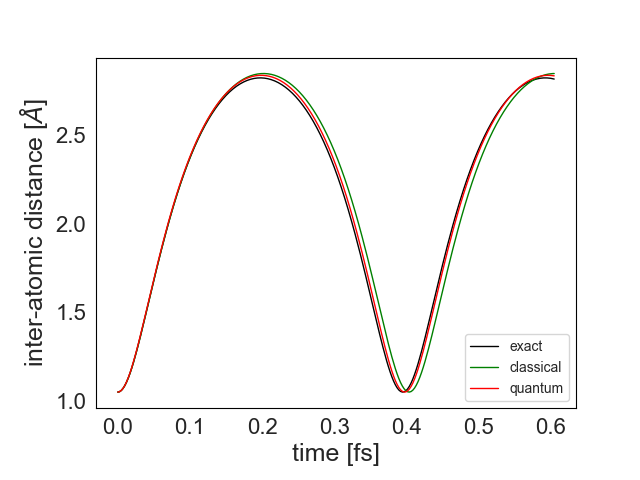} }%
    \qquad
    \subfloat[\centering Oscillation frequencies ]{\includegraphics[width=8.5cm]{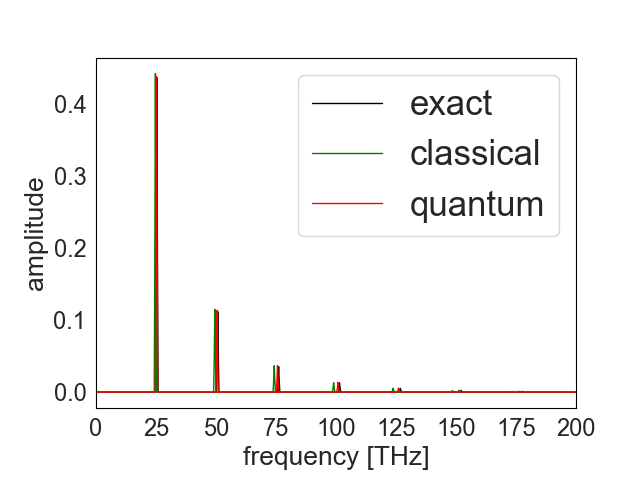} }%
    \caption{Molecular dynamics of a LiH molecule  with initial conditions $x_0=1.05$ $[\angstrom]$ and $v_0=0$ [at.u.]. The exact solution (blue) is compared to the classical NNP (green) and to the quantum NNP (red) ones.}
    \label{LiH_x_MD_1.05}
\end{figure}

\section{LiH -- Molecular Dynamics}\label{app:md}

The motivation behind both quantum and classical neural network potentials is ultimately to provide computationally effective access to highly accurate force fields to drive molecular dynamics. In this section, we provide a first demonstration of a quantum NNP used to simulate the oscillations of a LiH molecule around its equilibrium position. We use the Velocity Verlet algorithm~\cite{VV} to compute the time evolution of inter\hyp atomic distance and velocity, starting from some given initial conditions. At each point, the forces are predicted with the trained QNN presented in Sec.~\ref{sec:LiHqnn}. For comparison, we also present results obtained with the classical NNP introduced in Sec.~\ref{sec:LiHqnn} of the main text.

We start with the atoms at rest and close to each other, with an initial inter\hyp atomic distance of $1.05$ \AA, and let the system evolve according to Netwon's equations. The time evolution of the inter\hyp atomic distance is shown in Fig.~\ref{LiH_x_MD_1.05}a, where we compare the results obtained with classical and quantum NNPs to the exact solution obtained numerically. To assess the overall quality of the trajectories, we compute the frequency spectrum of the oscillations with a Fourier transform, which we apply to an evolution of $0.6$ fs, repeated 1000 times and passed through an Hamming window. As reported in Fig.~\ref{LiH_x_MD_1.05}b, the quantum and classical models can accurately reproduce the dominant frequencies.

\section{H$_3$O$^+$ -- Internal coordinates}\label{app:dihedral}
A hydronium molecule has $6 = 4\times 3 - 6$ degrees of freedom. To describe its configurations, we used 3 bond lengths (r$_{\text{OH,1}}$, r$_{\text{OH,2}}$ and r$_{\text{OH,3}}$), 2 angles ($\phi_{\text{H,1OH,2}}$ and $\phi_{\text{H,1OH,3}}$) and 1 dihedral angle (d$_{\text{OH,3H,2H,1}}$). We used the following definition for the dihedral angle with atoms $(ijkl)$:
\begin{align}
    d_{ijkl} &= \text{sign}(\chi)\arccos{\frac{(\vec{r}_{ij}\times \vec{r}_{kj})\cdot (\vec{r}_{kj}\times \vec{r}_{kl})}{|\vec{r}_{ij}\times \vec{r}_{kj}|\cdot |\vec{r}_{kj}\times \vec{r}_{kl}|}} \\
    \chi &= \vec{r}_{kj}\cdot(\vec{r}_{ij}\times \vec{r}_{kj}) \times (\vec{r}_{kj}\times \vec{r}_{kl}),
\end{align}
where $\vec{r}_{ij}$ is the distance vector between atom $i$ and $j$. The above definitions and the formulas for their gradients can be found in Ref.~\cite{internal}.

\end{document}